\documentclass[aps,groupedaddress,epsfig,showpacs,floats]{revtex4}
\usepackage{graphicx}

\begin{document}

\title{Influence of the stray field of magnetic dot on the nucleation of superconductivity in a disk}
\author{ D. S. Golubovi\'{c}\footnote{Dusan.Golubovic@fys.kuleuven.ac.be}, W. V. Pogosov, M. Morelle and V. V. Moshchalkov}
\affiliation{Nanoscale Superconductivity and Magnetism Group,
Laboratory for Solid State Physics and Magnetism , K. U. Leuven,
Celestijnenlaan 200 D, B-3001 Leuven, Belgium}

\begin{abstract}
We have investigated the nucleation of superconductivity in an Al
mesoscopic disk, with a magnetic dot on the top. The dot is
magnetized perpendicularly, and its magnetic field is
inhomogeneous. The Al disk and magnetic dot are separated by an
insulating layer, which ensures that there is only magnetic
interaction between them. This hybrid superconductor/ferromagnet
structure exhibits the maximum critical temperature for a finite
value of the perpendicular applied magnetic field, which is
parallel to the magnetization of magnetic dot. We have found a
good agreement between the experimental data for the
superconductor/normal metal phase boundary and the theoretical
predictions based on the Ginzburg-Landau theory.
\end{abstract}

\pacs{74.78.Na,74.25.Dw}
\maketitle

\section{Introduction}
{\it Hybrid} superconductor/ferromagnet structures have attracted
a lot of attention
\cite{victor,ja,pokro,rusi,misko,miskoa,mjvb,martin,mjvbb}. So
far, the experimental efforts have mainly been focused on
superconducting thin films with arrays of magnetic dots. Recently,
a superconducting disk with a perpendicularly magnetized dot was
fabricated and its superconducting $T_{c}(B)$ phase boundary
determined \cite{ja}. However, the nucleation of superconductivity
in this case was strongly affected by the proximity effect between
the disk and dot.

In this paper we have investigated the onset of superconductivity
in an Al disk with a perpendicularly magnetized Co/Pd magnetic dot
on the top. The dot is separated from the disk by an insulating
spacer layer, which ensures that there is no suppression of the
order parameter in the disk due to the proximity effect and that
the interaction between the disk and dot has only {\it magnetic}
character \cite{ja}.

The superconducting $T_{c}(B)$ phase boundary, obtained by
transport measurements, exhibits an asymmetry with respect to the
polarity of the applied magnetic field. The maximum critical
temperature, higher than the zero-field critical temperature, is
attained for a finite applied magnetic field which is parallel to
the magnetization of magnetic dot.

\section{Sample preparation and experimental technique }
The sample was prepared on a SiO$_{2}$ substrate by electron beam
lithography on PMMA950K and the co-polymer electron beam resists
in three phases. Each phase involves the patterning of a desired
structure, thermal evaporation of the material and ultrasonic
assisted lift-off procedure.

The contact pads and leads, as well as the alignment markers are
made up of $5\,$nm Cr and $30\,nm$ Au. The superconducting disk is
a $60\,$nm thick Al, whereas the magnetic dot consists of
$2.5\,nm$ Pd buffer layer and $10$ bilayers of $0.4\,$nm Co and
$1\,$nm Pd. A $10\,$nm thick Si spacer layer was evaporated before
the magnetic dot. A careful alignment procedure was needed to
position the dot at the centre of the disk.

\begin{figure}[htb]
\centering
\includegraphics*[width=8cm]{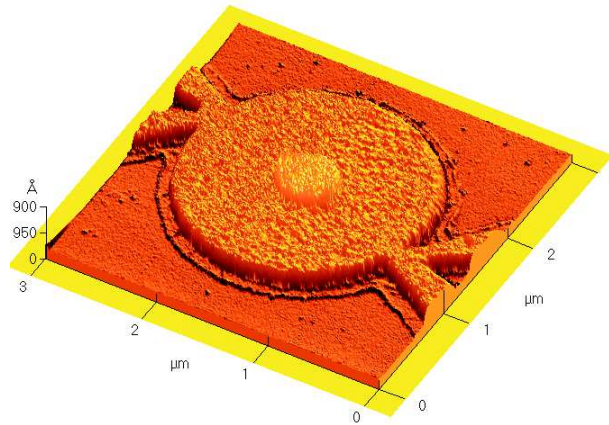}
\caption{An AFM topography image of the sample. \label{afm}}
\end{figure}

Fig. \ref{afm} presents an atomic force microscopy image of the
sample. The radius of the Al disk is $1.08\,{\rm \mu m}$, whereas
the dot has the radius of $270\,$nm. The electric contacts are
wedge-shaped, with the opening angle of $15\,^{\circ}$, as these
have proved to be the least invasive for transport measurements of
superconducting nanostructures \cite{proka}.

The magnetic properties of patterned Co/Pd structures were
thoroughly investigated in Ref. \cite{martin}. We have determined
the magnetic properties of the Co/Pd multilayer at room
temperature from the magneto-optical Kerr measurements of the
co-evaporated plane film. These data have confirmed that the Co/Pd
multilayer has a perpendicular anisotropy, with a complete
remanence and coercive field of approximately $70\,$mT. Prior to
the measurements the sample was magnetized perpendicularly in the
magnetic field of $300\,$mT. As the applied magnetic fields in the
experiment never exceeded $30\,$mT, we have assumed that the
magnetization of the dot remains unaltered during the
measurements.

The onset of superconductivity in the structure was studied by
measuring the superconducting $T_{c}(B)$ phase boundary. The phase
boundary was found resistively, from four-point transport
measurements, in a cryogenic setup at temperatures down to
$1.11\,$K, with the temperature and field resolution of $0.5\,$mK
and $5\,{\rm \mu T}$, respectively. The transport current with the
effective value of $100\,$nA and frequency $27.7\,$Hz was used,
whilst the signal-to-noise ratio was being improved by a lock-in
amplifier.

The resistance of the structure at room temperature is $4.85\,{\rm
\Omega}$, the low temperature resistance is $R_{n}=2.8\,{\rm
\Omega}$, whereas the maximum critical temperature is
$T_{cm}=1.421\,$K.

\section{Experimental results}

\begin{figure}[htb]
\centering
\includegraphics*[width=8.5cm]{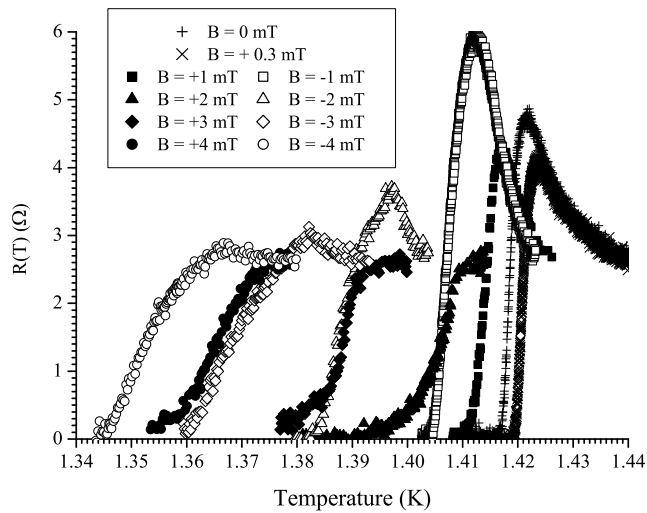}
\caption{$R(T)$ transition in different applied magnetic fields.
Filled symbols indicate the transitions in the applied fields
parallel to the magnetization of the dot, whereas open symbols
present the transition in the antiparallel magnetic fields.
\label{rt}}
\end{figure}

Fig. \ref{rt} presents resistive transitions of the structure in a
constant applied magnetic field. We have taken parallel magnetic
fields as positive and antiparallel magnetic fields as negative,
and will be using this convention throughout the paper.

Typically, superconducting structures have resistive transitions
that are symmetric with respect to the polarity of an applied
magnetic field. Likewise, the maximum transition temperature is
achieved in zero applied field. The transitions in Fig. \ref{rt}
are strongly asymmetric with respect to the polarity of an applied
field. The critical temperature, defined conventionally as the
temperature at which the resistance is $R_{n}/2$, in the applied
magnetic field of $+2\,$mT is equal to the transition temperature
in $-1\,$mT. This dependence of the resistive transitions on the
field polarity is reproduced for higher fields, as well. More
importantly, the critical temperature in zero applied field is not
the maximum critical temperature of the structure. The structure
attains the maximum critical temperature when exposed to the
magnetic field of $+0.3\,mT$. The difference between the maximum
 and zero-field critical temperature
 is approximately $2.5\,$mK.

The transitions for the applied fields $-1 < B_{a}<1\,{\rm [mT]}$
have considerable overshoots in resistance, with respect to the
normal state. This phenomenon is related to the formation of
normal/superconducting (NS) interfaces and nonequilibrium charge
imbalance effects \cite{vital,vit,proka}. The critical temperature
of the mesoscopic contacts is slightly higher than the critical
temperature of the disk \cite{proka}. For this reason,
superconductivity nucleates nonuniformly, thus  giving rise to the
formation of the NS interfaces. Due to a finite local magnetic
induction in the disk generated by the dot, the difference in the
critical temperatures of the disk and the mesoscopic contacts is
greater than for a disk without a magnetic dot. As a positive
applied magnetic field increases, the total magnetic induction in
the disk around the magnetic dot decreases, thereby effectively
reducing the difference in the critical temperatures of the disk
and mesoscopic contacts. For this reason, as the positive applied
field increases the amplitude of the resistance anomaly decreases.
On the other hand, a negative applied field gives rise to a
greater difference in the local critical temperatures, which is
responsible for more pronounced peaks. When increasing the
negative applied field, the contribution of the stray field to the
total field becomes less significant and the typical amplitudes of
the overshoots are recovered \cite{victor,proka}.

\begin{figure}[htb]
\centering
\includegraphics*[width=8.5cm]{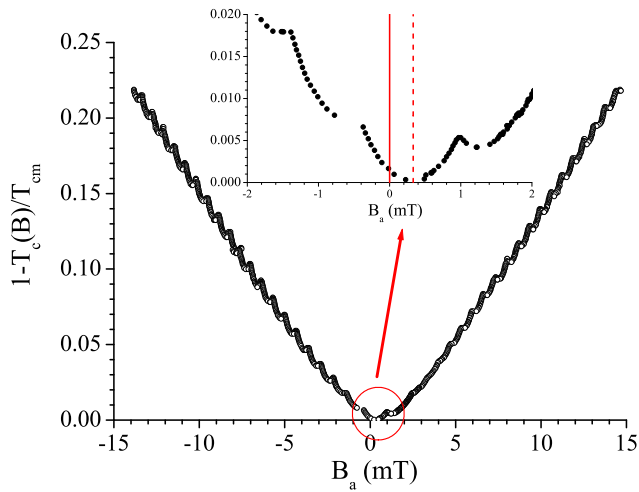}
\caption{ The experimentally obtained superconducting phase
boundary plotted as $1-T_{c}(B)/T_{cm}$ versus the applied field
$B_{a}$. $T_{cm}$ is the maximum critical temperature. The inset
shows the phase boundary around $T_{c0}$. \label{fig4}}
\end{figure}

Fig. \ref{fig4} shows the experimental phase boundary in the plane
of the normalized critical temperature $1-T_{c}(B)/T_{cm}$ and the
applied magnetic field, with $T_{cm}$ being the maximum measured
critical temperature. The inset displays the phase boundary around
the zero-field critical temperature $T_{c0}$. The phase boundary,
except for the low fields, has a {\it linear} background, which is
the hallmark of the nucleation process in the disk \cite{victor}.
In addition, the quasiperiodicity of the phase boundary is
entirely consistent with what has been obtained for a mesoscopic
superconducting disk without the dot \cite{vb}. When there is a
pronounced proximity effect between the superconducting disk and
magnetic dot, the order parameter is suppressed below the dot and
the disk can be approximated with a superconducting loop with a
finite width \cite {ja}. In this case, the superconducting
$T_{c}(B)$ phase boundary has a crossover in the background - from
a parabolic for low fields to a linear for high fields \cite{vb}.
Therefore, the presence of a pronounced proximity effect between
the dot and the disk can confidently be ruled out, and specific
features of the phase boundary can be attributed to the magnetic
interaction between the disk and magnetic dot.

The most striking feature of the phase boundary is its asymmetry.
The minimum in the phase boundary, which corresponds to the
maximum critical temperature of the structure, has been obtained
for a finite applied magnetic field which is parallel to the
magnetization of magnetic dot.

\section{Theoretical results}

 The experimental data have been analysed by using the
Ginzburg-Landau (GL) theory. Near the transition from the normal
to the superconducting state the total magnetic field in the disk
is equal to the sum of the stray field of the magnetic dot and a
uniform externally applied field. The stray field we have used in
the analysis, was obtained by magnetostatic calculations. For
details, we refer to \cite{ja,misko}.

As the sample in our experiment is thinner than the coherence
length $\xi (T)$, the order parameter is constant along $z$ axis
and the magnetic field in the GL equations can be averaged out
over the thickness of the disk. Thus, the problem is reduced to
the 2D case. The axial symmetry of the sample allows us to further
reduce the dimensionality of the problem to the 1D case, since
near the phase boundary the modulus of the order parameter is
axially symmetric. Using the cylindrical coordinate system with
coordinates $r$, $\varphi $, $z$, the dimensionless order
parameter $\psi(r,\phi)$ can be expressed as
\begin{equation}
\psi(r,\phi)=f(r)\, {\rm exp}(-iL\phi) \\\ ,  \label{ord}
\label{eq1}
\end{equation}
where $f(r)$ is the modulus of the order parameter and $L$ stands
for the winding number (vorticity). Instead of solving the GL
equations, which are nonlinear with respect to $f(r)$, we have
applied a variational procedure, similar to what has been used in
Ref. \cite{pogosov}. The trial function used for the modulus of
the order parameter is
\begin{eqnarray}
f(r) = p_{1}\cdot {\rm exp} \left(-q {r^{2}\over{R^{2}}} \right)
\cdot \left(\left({r\over{R}}\right)^{L}+p_{2}\left({r\over{R}}
\right)^{L+1}+p_{3}\left({r\over {R}}\right)^{L+2}+p_{4}\left(
{r\over{R}} \right)^{L+3} \right)  . \label{eq2}
\end{eqnarray}
where $p_{1}$, $p_{2}$, $p_{3}$ and $p_{4}$ are the variational
parameters, $R$ is the radius of the disk, whereas $q$ is found
from the vacuum boundary condition for the order parameter
($f'(R)=0$)
\begin{equation}
q={L+p_{2}(L+1)+p_{3}(L+2)+p_{4}(L+3)
\over{2(1+p_{2}+p_{3}+p_{4})}} .
\end{equation}

Using Eqs. (\ref{eq1}) and (\ref{eq2}) the GL energy is found as a
function of the variational parameters (see Ref. \cite{pogosov}).
The values of the variational parameters are calculated by
minimizing the GL energy. Comparing the energies of states with
different $L$ the superconducting $T_{c}(B)$ phase boundary of the
disk is found.

\begin{figure}[htb]
\centering
\includegraphics*[width=8.5cm]{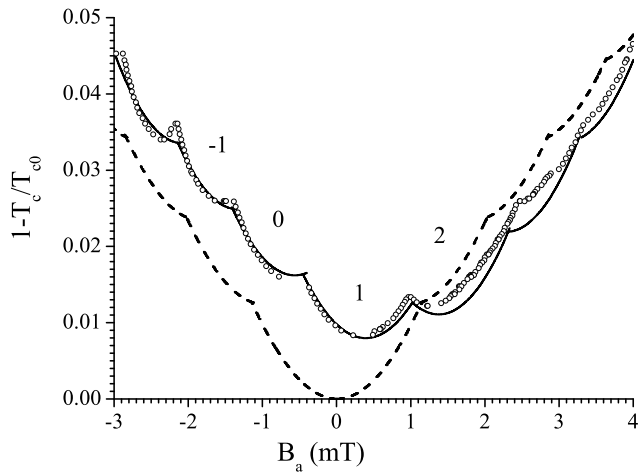}
\caption{The experimental data and the corresponding theoretical
phase boundary of the superconducting disk with magnetic dot
(solid line), as well as the theoretical phase boundary of an
identical superconducting disk without magnetic dot (dashed line).
The temperatures are normalized to the zero-field critical
temperature of the disk without magnetic dot. The numbers indicate
the vorticity of the hybrid structure. \label{ter}}
\end{figure}

Fig. \ref{ter} displays the theoretical fit of the experimental
phase boundary in the low field regime, along with the theoretical
phase boundary of an identical superconducting disk without
magnetic dot (dashed line). The critical temperatures are
normalized to the zero-field critical temperature of the
superconducting disk without magnetic dot, as obtained by the
calculations. The best agreement between the theory and the
experiment has been found for $\xi (T=0)=90\,$nm, which is
consistent with the value of $\xi(T=0)$ obtained in Ref.
\cite{victor} for mesoscopic Al superconductors.
Each cusp in the phase boundary corresponds to
the transition between the states with different vorticities.
According to our results, there is one vortex in the disk in the
absence of the external magnetic field.

The superconducting phase boundary strongly depends upon the
polarity of an external magnetic field.  The magnetization of the
dot $m$ has been chosen to provide the best qualitative and
quantitative agreement between the theory and the experiment in
the vicinity of $T_{c0}$. The direction of the shift of $T_{c}(B)$
phase boundary near $T_{c0}$, for a fixed orientation of the
magnetization $m$, depends upon the intensity of the stray field
of magnetic dot. The shift can come about as a result of the
cancellation of the total flux generated by the magnetic dot, or
due to a change in the kinetic energy of the superconducting
condensate in the disk, accompanied by a switch in the vorticity
by one. The former shifts the phase boundary in the direction
opposite to the magnetization of the dot and the maximum critical
temperature is observed for a finite {\it negative} applied field,
whereas the latter provides that the maximum critical temperature
is achieved for a finite applied field parallel to the
magnetization of the dot, that is for a finite {\it positive}
field. Which of these competing effects prevails strongly depends
upon the intensity of magnetization of magnetic dot, as well as
upon the parameters of the superconducting structure. In our case
the shift is positive.

\begin{figure}[htb]
\centering
\includegraphics*[width=5.8cm]{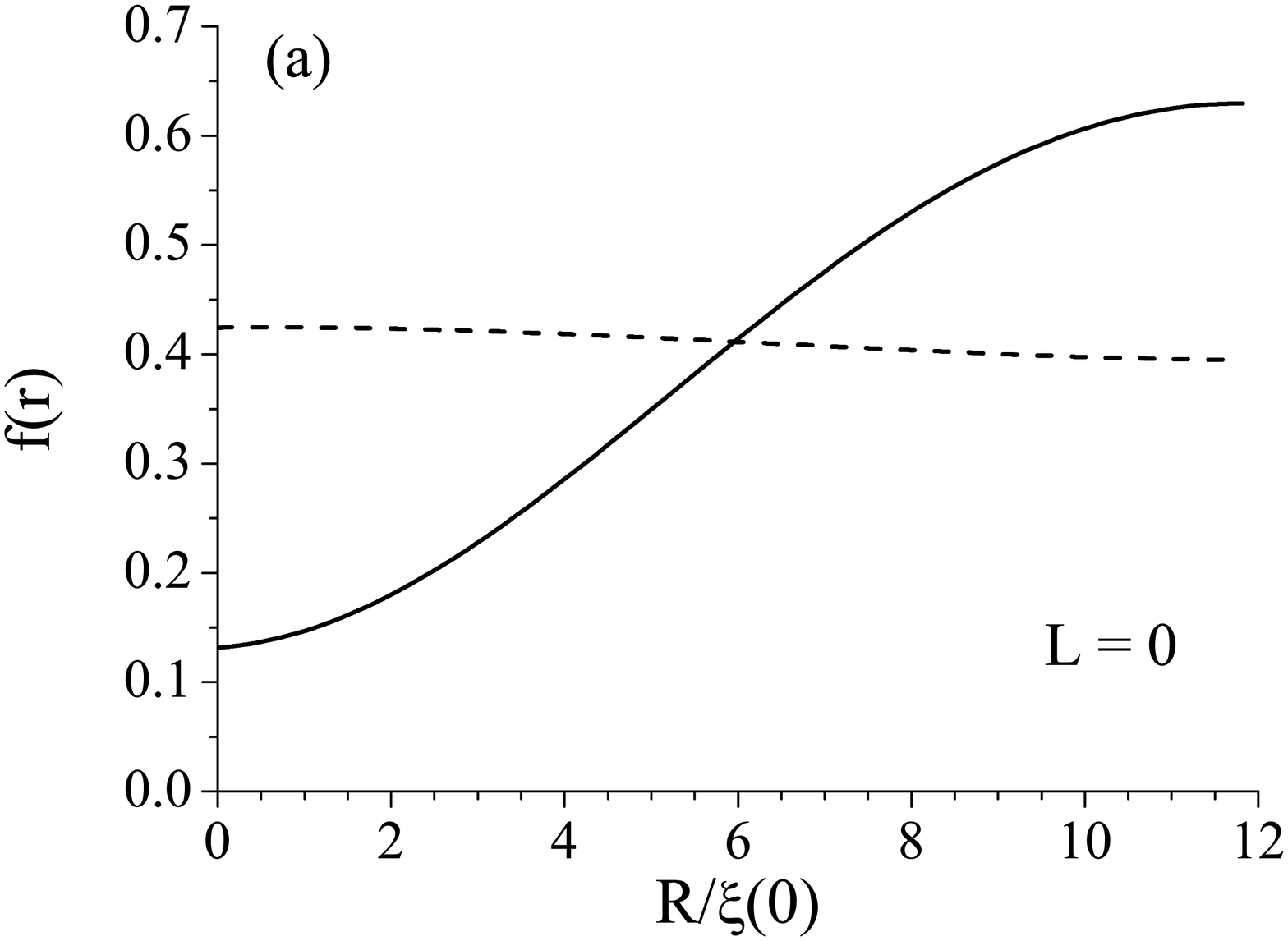}
\includegraphics*[width=5.8cm]{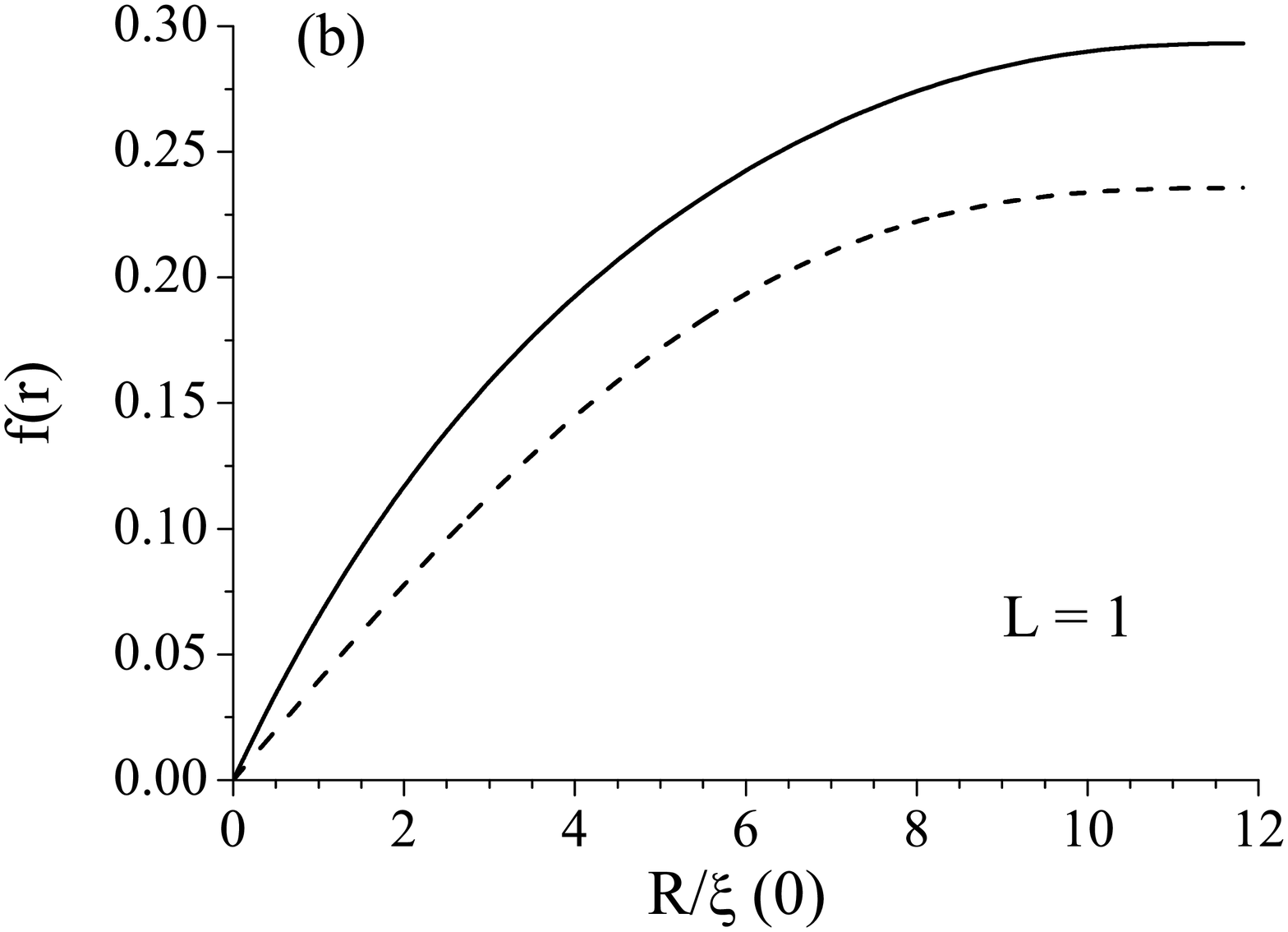}
\caption{ The moduli of the order parameter in (a) the Meissner
state and (b) vortex state with the vorticity $L=1$. The solid
line describes the superconducting disk with magnetic dot, whereas
the dashed line presents the disk without magnetic dot. In the
Meissner state the temperatures are $T/T_{c0}=0.987$ and
$T/T_{c0}=0.995$ for the disk with and without magnetic dot,
respectively. For the vortex state $L=1$ the respective
temperatures are $T/T_{c0}=0.990$ and $T/T_{c0}=0.995$. Here
$T_{c0}$ is the zero-field temperature of the plane
superconducting disk. \label{psi}}
\end{figure}

In addition to the shift along the $B$-axis, the phase boundary of
the hybrid structure is also shifted along the $T$-axis. The shift
along the $T$-axis is caused by a difference in the maximum
critical temperature of the disk with and without the dot. As the
stray field of the dot is spatially inhomogeneous, it cannot be
cancelled out by an externally applied homogeneous magnetic field.
For this reason, there is always a finite flux through a disk with
a magnetic dot and its maximum critical temperature is reduced
compared to the disk without a magnetic dot. Even though the
maximum critical temperature of the hybrid structure is less than
1$\, \%$ lower than the maximum critical temperature of the plane
disk, this difference is sufficient to modify  the phase boundary.
If the additional homogeneous magnetic field were applied, there
would be no differences in maximum critical temperature, and,
consequently, the $T_{c}(B)$ phase boundary would not be
substantially modified along the $T$-axis. We can, therefore,
conclude that the effect of the {\it inhomogeneous} magnetic field
on the nucleation of superconductivity is {\it twofold}: it shifts
the phase boundary along the $B$-axis, as well as distorts the
phase boundary along the $T$-axis, altering the values at which
the structure switches between different vorticities.

The stray field of the magnetic may change the typical spatial
profile of the superconducting condensate density within the disk.
Fig. \ref{psi}(a) and \ref{psi}(b) present the moduli of the order
parameter for the disk with magnetic dot (solid line) and disk
without magnetic dot (dashed line). Fig. \ref{psi}(a) shows the
Meissner state, whereas Fig. \ref{psi}(b) shows the state with the
vorticity $L=1$. In the Meissner state, for the plane
superconducting disk, the modulus of the order parameter has a
maximum at the centre of the disk. On the other hand, the
amplitude of the order parameter is reduced below the magnetic
dot, where the intensity of the stay field is the highest, and
exhibits a minimum at the centre of the disk. The spatial profile
of the order parameter is not strongly affected by the stray field
in the vortex state, because the vortex core, where the order
parameter is equal to zero, is in the region below the dot.

In conclusion, we have fabricated a mesoscopic superconducting
disk made up of Al with a perpendicularly magnetized magnetic dot
on the top. The superconducting properties of the system have been
investigated by measuring the superconducting/normal state phase
boundary $T_{c}(B)$. It has been demonstrated that the phase
boundary is asymmetric with respect to the direction of the
applied field. The maximum critical temperature has been attained
for a finite value of the applied magnetic field, which is
oriented parallel to the magnetization of magnetic dot. The
experimental data are in a good agreement with the theoretical
results. It has also been shown that the inhomogeneity of the
stray field gives rise to a modification of the $T_{c}(B)$ phase
boundary along the $T$-axis, which would not be present if an
additional magnetic field were homogeneous.

\section{Acknowledgements}
The authors would like to thank G. Rens for AFM measurements. This
work has been supported by the Belgian IUAP, the Flemish FWO and
the Research Fund K. U. Leuven GOA/2004/02 programmes, as well as
by the ESF programme "VORTEX". W. V. P. acknowledges the support
from the Research Council of the K.U. Leuven and DWTC.

\end{document}